\documentclass[prl, showpacs, twocolumn, floatfix]{revtex4}
\usepackage{graphicx}
\usepackage{amsmath, amsfonts, amssymb, bm}
\begin{document}
\title{Enhanced lifetime of positronium atoms via collective radiative effects}

\author{Ni \surname{Cui}}
\email{ni.cui@mpi-hd.mpg.de}
\author{Mihai \surname{Macovei}}
\email{mihai.macovei@mpi-hd.mpg.de}
\author{Karen Z. \surname{Hatsagortsyan}}
\email{k.hatsagortsyan@mpi-hd.mpg.de}
\author{Christoph H. \surname{Keitel}}
\email{keitel@mpi-hd.mpg.de}

\affiliation{Max-Planck-Institut f\"{u}r Kernphysik, Saupfercheckweg
1, D-69117 Heidelberg, Germany}
\date{\today}

\begin{abstract}
A method is proposed to manipulate the annihilation dynamics of a dense gas of positronium atoms employing
superradiance and subradiance regimes of the cooperative spontaneous emission of the system. The annihilation dynamics  is controlled by the gas density and by the intensity of the driving strong resonant laser field. In particular, the method allows to increase the annihilation lifetime of an ensemble of positronium atoms more than hundred times by trapping the atoms in the excited state via collective radiative effects in the resonant laser and cavity fields.
\end{abstract}

\maketitle
Positronium (Ps) is a hydrogen-like atom comprised of an electron and its antiparticle
- the positron~\cite{Rich}. As a purely leptonic system with a small mass and long de Broglie
wavelength, Ps provides an unique opportunity to form a Ps Bose-Einstein condensate (BEC) at
relatively high temperatures~\cite{Mills-1994,Mills-2010,Cassidy}. Furthermore, stimulated annihilation from a Ps BEC may set up a route towards a $\gamma$-ray laser~\cite{Mills-2002,gamma-laser,gamma-laser2}. In another front, Ps provides a way for antihydrogen formation via the charge transfer process $\rm Ps+\bar{p}\rightarrow e^-+\bar{H}$ \cite{anti-H}.
Recently, the techniques for production, accumulation, control and manipulation of positronium atoms have been essentially advanced \cite{Mills}. This allowed, for example, the observation of molecular Ps when injecting an intense positron beam  on a thin film of porous silica \cite{Mills_Nature}.

Different from ordinary atoms, the Ps atom is unstable. The singlet Ps (para-Ps) annihilates mainly by
two $\gamma$-photon emission with a short lifetime $\tau_s=1.25\times 10^{-10} \rm s$ whereas the triplet Ps
(ortho-Ps) annihilates through three $\gamma$-photon emission with a longer lifetime
$\tau_t=1.4\times 10^{-7} \rm s$~\cite{Rich}. Such short lifetimes give rise to evident difficulties in
accumulating Ps atoms and achieving Ps BEC.
Extending the Ps lifetime will considerably increase the feasibility
for Ps BEC, even with a singlet Ps. The Ps atoms are produced predominantly in the $n=1$ ground
state; only about one in $10^3-10^4$ atoms is found in the $n=2$ upper state~\cite{Mills-1975}.  For states
with a high angular momentum, the annihilation rate is much smaller than the rate for the respective 
$s$-state, and the lifetime is considerably larger. To extend the Ps lifetime, one may keep the atoms away
from the $s$-state.
The optical pumping has been proposed as a method for populating the excited states of Ps atoms and for increasing the Ps lifetime.
In particular, Pazdzerskii~\cite{Pazdzerskii}, Faisal~\cite{Faisal} and
Mittleman~\cite{Mittleman-1986} have proposed to use a resonant laser field 
to excite Ps out of the ground $s$-state to a higher
$p$-state. With a single resonant field the best result that can be achieved is the doubling of the Ps lifetime.
When using two resonant driving laser fields which couple three states of Ps in a V-type system \cite{Mittleman},  a better result via coherent population trapping  scheme can be achieved,
increasing the annihilation lifetime by about a factor of 2 for the triplet case and 20 for the singlet case.
More efficient excitation of Ps atoms are shown in \cite{Lambropoulos} at the expense of applying stronger laser fields (up to intensities of $10^{13}$ W/cm$^2$) and at a high ionization background which is hardly applicable in the case of the BEC formation.
The first experimental observation of resonant laser excitation of high-$n$ states in Ps in a magnetic environment has been carried out by Ziock et al. \cite{Ziock}.

In this Letter, we investigate the role of cooperative radiative processes in a dense gas of Ps atoms on the population dynamics and its influence on the annihilation evolution of the ensemble. Employing superradiant spontaneous emission, controlled by the density of the gas, and combining it with a resonant driving laser field, a population trapping scheme is developed which can significantly enhance the lifetime of the Ps ensemble. Moreover, the characteristics of the trapping scheme is shown to be essentially improved when the cavity accommodating the Ps gas constitutes a radiation cavity.

\begin{figure}[t]
\includegraphics[width=8cm]{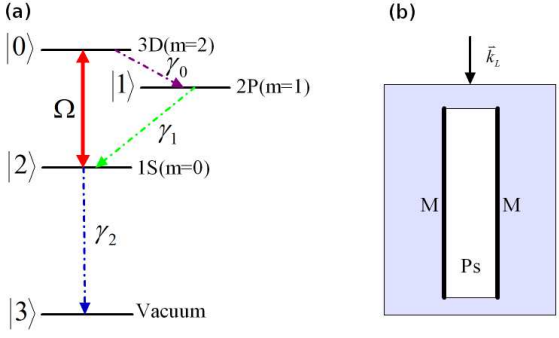}
\caption{\label{fig-1}(color online) (a) The energy levels of a Ps atom.
The coherent laser drives the $|0\rangle \leftrightarrow |2\rangle$ atomic transition with a Rabi frequency
$\Omega$. $\gamma_{0}$ and $\gamma_{1}$ are the single-atom spontaneous decay rates on transitions
$|0\rangle \to |1\rangle$ and $|1\rangle \to |2\rangle$, respectively, whereas $\gamma_{2}$ describes the
annihilation decay rate. (b) The cross-section of the cavity accumulating the Ps atoms. A flux of positrons impinges (transverse to the plane of the figure) on a porous silica film  and the created Ps atoms are accumulated in a hole (cavity) inside the film. Mirrors M are mounted in the hole forming a radiation cavity which is resonat to the  $|0\rangle \to |1\rangle$ atomic transition. $\vec{k}_L$ is the wave-vector of the strong laser field coupling the states $|0\rangle$ and $|2\rangle$.}
\end{figure}

In a dense gas of Ps atoms, cooperative effects can significantly influence the spontaneous emission processes if the inter-particle distances are less than the involved radiation wavelengths. The annihilation produces $\gamma$-rays with a wavelength $\lambda\sim \lambda_C$, where $\lambda_C =3.86\times 10^{-11}$ cm is the Compton wavelength. The cooperative effects play a  role in the case of annihilation only if the density of the Ps gas is larger than the characteristic density $\rho_c\sim \lambda^{-3}\sim 10^{31}$ cm$^{-3}$. Since the achievable densities for the Ps accumulation are much smaller, the superradiant enhancement of the annihilation is not realistic. However, the  wavelengths of bound-bound transitions are rather large which allows in principle for collective radiative effects to enter into play. In particular, the transition from $n=2$ to $n=1$ states corresponds to a wavelength of $\lambda_{12}\sim 0.23\,\mu$m ($\rho_c\sim 10^{14}$ cm$^{-3}$), while the transition from $n=3$ to $n=2$ to $\lambda_{01}\sim 1.2\,\mu$m ($\rho_c\sim 8\times 10^{12}$ cm$^{-3}$), see Fig.~(\ref{fig-1}a). Such densities are within the reach for modern Ps accumulation techniques \cite{Cassidy_2005}. Therefore, the selective atomic spontaneous transitions can be cooperatively enhanced by means of adjusting the density of the gas. This opens a way for controlling the population of Ps excited states and, via coherently coupling the bound-bound and bound-free transitions, for controlling the annihilation dynamics. 

We consider an ensemble of singlet Ps atoms with a high density,  allowing a significant interaction between atoms via the common vacuum field. The model of our system is characterized by $N$ identical nonoverlapping three-state Ps atoms, consisting of 3D state $|0\rangle$, 2P state $|1\rangle$, 1S state $|2\rangle$ and the vacuum state $|3\rangle$, as shown in Fig.~\ref{fig-1}(a). Our model includes resonant coupling of the atomic transition between the states 3D and 2P
by a coherent laser field. Other dipole-allowed transitions $3D \leftrightarrow 2P$ and $2P \leftrightarrow 1S$ are 
coupled via vacuum modes with dipole moments $d_{01}$ and $d_{12}$, respectively. The Ps atoms in the singlet 1S ground state  annihilates through two $\gamma$-photon emission and the vacuum state $|3\rangle$ is introduced to account for the annihilation channel (see Fig.~\ref{fig-1}a). 
In the dipole approximation, the Hamiltonian of the total system reads:
\begin{eqnarray}
H=H_A+H_F+H_{AF}+H_L+H_D, \label{Hm}
\end{eqnarray}
where $H_A$ and $H_F$ describe the free time evolution of the atoms and the vacuum radiation field, respectively: 
\begin{eqnarray}
H_A=\sum_{\alpha=0}^{2}\sum_{j=1}^{N}\hbar\omega_\alpha^{(j)}S_{\alpha\alpha}^{(j)}, \label{Ha}
~~~H_F=\sum_{\vec{k}s}{\hbar\omega_{\vec{k}s}\hat{a}^+_{\vec{k}s}}\hat{a}_{\vec{k}s}.\label{Hf}
\end{eqnarray}
The interaction between the Ps atoms and the quantized vacuum radiation field is described by
\begin{eqnarray}
H_{AF}&=&i\sum_{j=1}^{N}\sum_{\vec{k}s}(\vec{d}_{01}\cdot \vec{g}_{\vec{k}s})(\hat{a}^{+}_{\vec{k}s} S_{10}^{(j)}e^{-i\vec{k}\cdot\vec{r}_j}-\text{H.c.})\nonumber\\
&+&i\sum_{j=1}^{N}\sum_{\vec{k}s}(\vec{d}_{12}\cdot \vec{g}_{\vec{k}s})(\hat{a}^{+}_{\vec{k}s}
S_{21}^{(j)}e^{-i\vec{k}\cdot\vec{r}_j}-\text{H.c.}),
\label{Haf}
\end{eqnarray}
where $\vec{g}_{\vec{k}s}=\epsilon_{\vec{k}s}\sqrt{2\pi\hbar\omega_{\vec{k}s}/V}$ and $\hbar\omega^{(j)}_{\alpha}$ with $\alpha\in\{0,1,2\}$ are the energies of the state $|\alpha\rangle$ of the $j$-th atom. $S^{(j)}_{\alpha\beta}=|\alpha\rangle_{jj}\langle\beta|$ with $\{\alpha,\beta\} \in\{0,1,2,3\}$ represents the 
population in the state $|\alpha\rangle$ of the $j$-th atom if $\alpha=\beta$ or the transition operator from 
$|\beta\rangle$ to $|\alpha\rangle$ of the $j$-th atom if $\alpha\neq\beta$ while obeying the commutation relations $\left[S^{(j)}_{\alpha\beta},S^{(l)}_{\beta'\alpha'}\right]=\delta_{jl}\left(\delta_{\beta\beta'}S^{(j)}_{\alpha\alpha'}- \delta_{\alpha\alpha'}S^{(j)}_{\beta'\beta} \right)$. Furthermore,  
$\hat{a}_{\vec{k}s}~\left(\hat{a}^{\dagger}_{\vec{k}s}\right)$ is the annihilation (creation) operator of the photon
with the wave vector $\vec{k}$ and $\epsilon_{\vec{k}s}$ represents the unit polarization vector with $s\in\{1,2\}$ and 
frequency $\omega_{k}$. $\vec{r}_j$ is the position of the $j$-th Ps atom.
$H_L$ represents the interaction Hamiltonian of the Ps atoms with the resonant laser
field
\begin{eqnarray}
H_L=\sum_{j=1}^{N}\hbar\Omega(S_{02}^{(j)} e^{-i\omega_{02} t+2i\vec{k}_L \cdot \vec{r}_j} + \text{H.c.}), \label{Hl}
\end{eqnarray}
with $\Omega$ and $\vec{k}_L$ being the Rabi frequency and the wave-vector of the laser field, respectively, and $\omega_{\alpha\beta}=\omega_\alpha-\omega_\beta$ is the 
transition frequency of the $|\beta\rangle \leftrightarrow |\alpha\rangle$ transition.
$H_D$ describes the annihilation of Ps atoms
from the ground state into high-energy photons~\cite{Mittleman}.
It is assumed that the interparticle separations among the Ps atoms ($r_{jl}=|\vec{r}_j-\vec{r}_l|$) 
do not exceed the corresponding $3D \leftrightarrow 2P$ and
$2P \leftrightarrow 1S$ transition wavelengths. Then, in the usual Born-Markov, mean field and rotating-wave
approximation the Ps quantum dynamics is governed by the following master equation:
\begin{eqnarray}
\dot{\rho}(t) &+& i\Omega\sum_{j=1}^{N} \left[ S_{02}^{(j)} +S_{20}^{(j)},\rho \right]=
-\sum_{j,l=1}^{N}\{ \gamma_{jl}^{(0)}[S_{01}^{(j)},S_{10}^{(l)}\rho] \nonumber \\
&+& \gamma_{jl}^{(1)}[S_{12}^{(j)},S_{21}^{(l)}\rho]\}
-\gamma_{2}\sum_{j=1}^{N} [S_{23}^{(j)},S_{32}^{(j)}\rho]  + \text{H.c.},\label{meq}
\end{eqnarray}
where $\gamma_{jl}^{(i)}\equiv\gamma_{i}\left[\aleph_{jl}^{(i)} + i\Omega_{jl}^{(i)}\right]~(i\in\{0,1\})$
are the collective parameters \cite{Agarwal,Ficek,Macovei,Gross}, with $\aleph_{jl}^{(i)}$ and $\Omega_{jl}^{(i)}$ 
describing the mutual interactions among atomic pairs (for the definition see Eqs.~(2.52-2.54) in \cite{Macovei}), while $\gamma_{0}=2d_{01}^2\omega_{01}^3/(3\hbar c^3)=3.2\times 10^7~\rm s^{-1}$ and $\gamma_{1}=2d_{12}^2 \omega_{12}^3/(3\hbar c^3)=3.1\times 10^8~\rm s^{-1}$ are the single-atom spontaneous decay rates from $3D \to 2P$ and $2P  \to 1S$ \cite{Rich}, respectively, and $\gamma_2=1/\tau_s$ is the annihilation rate of the
singlet ground state of a Ps atom.

\begin{figure}[t]
\includegraphics[width=7.0cm]{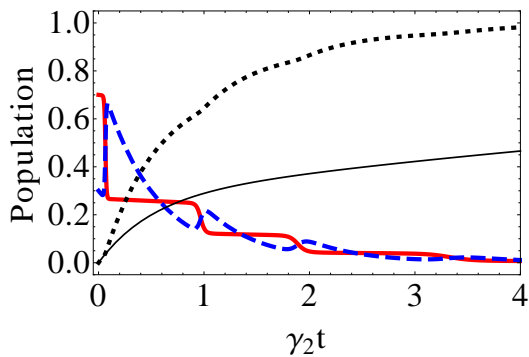}
\caption{\label{fig-2}(color online) The population of the $|1S\rangle$ ground
state (blue dashed curve) and the excited $|2P\rangle$ state (red thick solid curve) as a
function of time when initially 70$\%$ of atoms are prepared in the excited
$|2P\rangle$ state and the rest in the $|1S\rangle$ ground state. The annihilation probability for the ensemble (black dotted line) and for the independent Ps atoms (thin solid black curve) are also shown. Other parameters are: $N=10^{6}$, $\gamma_1/\gamma_2=1/25.8$
and $\mu_1=0.006$.}
\end{figure}

First, we investigate how the cooperative spontaneous transitions influence the annihilation dynamics of the dense ensemble of Ps atoms. We begin with the case of no driving resonant laser field ($\Omega=0$) and prepare the initial Ps ensemble in the excited $2P$ state. The decay $2P \rightarrow 1S$ occurs due to cooperative spontaneous emission.
There will be no population in the $3D$ state such that the first term in the r.h.s. and the second one in the l.h.s. of Eq.~(\ref{meq}) yield no contribution. In this situation, the Ps atom in the ensemble represents a two-level decaying (annihilating) system. The annihilation dynamics is investigated via numerical  integration of the master equation Eq.~(\ref{meq}). The equations for the populations are governed by the following parameters: the number of collectively interacting Ps atoms $N$,  the geometrical factor $\mu_1$  ($\mu_{1} \sim \lambda^{2}_{12}/d^{2}$, when the ensemble of Ps atoms has a form of a cylinder with a diameter $d$ \cite{Gross}) and decay rates $\gamma_{1,2}$. We choose $N=10^6$ 
such that $70\%$ are prepared in the excited
$2P$ state while the rest resides in the $1S$ ground state and $\mu_1=0.006$.
The evolution of populations in the $1S$ and $2P$ 
states as well as of the annihilation process are represented in Fig.~\ref{fig-2}.
For a comparison, the annihilation process for an independent Ps atom as a function of time is also shown in Fig.~\ref{fig-2}. The annihilation of the ensemble is altered due to collective spontaneous emission.
For an ensemble of $N$ initially excited two-level atoms the constructive interatomic interferences induce the atoms to radiate cooperatively with the probability 
proportional to $N^2$ and a quantum dynamics $N$ times faster than for an independent atom - an effect known as superradiance~\cite{Dicke,Gross}. 
The atoms in the excited $2P$ state suddenly decay to the $1S$ state in a short time at the very beginning which 
induces a sudden burst in the $1S$ state population.
After the superradiance burst, a subradiant state is formed and the population of the $2P$ state reaches a constant value which is different from zero for a while and there is no cooperative emission during this time.
In fact, it is well-known that in the (antisymmetrical) subradiant state  \cite{Dicke}, a pair of two-level radiators is unable to radiate  which is due to destructive interference when the atoms are not all de-excited~\cite{Crubellier,Pavolini}.
In contrast to the ordinary atoms, Ps in the $1S$ ground state will annihilate with an annihilation rate much larger than the radiative rate of the $2P\rightarrow 1S$ transition ($\gamma_2 \gg \gamma_1$). The annihilation effect breaks the spontaneously created subradiant state and the second superradiance burst appears. Because of the competition between the increase of population in the $1S$ state and the annihilation, the system reaches the second subradiant state, which is later destroyed by the annihilation again and followed by the third superradiant emission. The same processes will occur until there is no population in the $2P$ state anymore. The time delay between superradiant transitions, $T_d$, is determined by the annihilation rate: $T_d~\approx~\gamma_{2}^{-1}$, see Fig.~(\ref{fig-2}). The peaks in the $1S$ state population create modulation of the annihilation yield, see the dotted curve in Fig.~\ref{fig-2}, with a periodicity of $T_d$. The latter can be controlled 
by modifying the Ps density, the geometrical factor $\mu_1$, and by the initial population preparation.
Thus, there is a direct link between the collective effects and annihilation dynamics which, in particular, can be used to enhance the Ps lifetime. 

\begin{figure}[b]
\includegraphics[width=4.35cm,height=3.3cm]{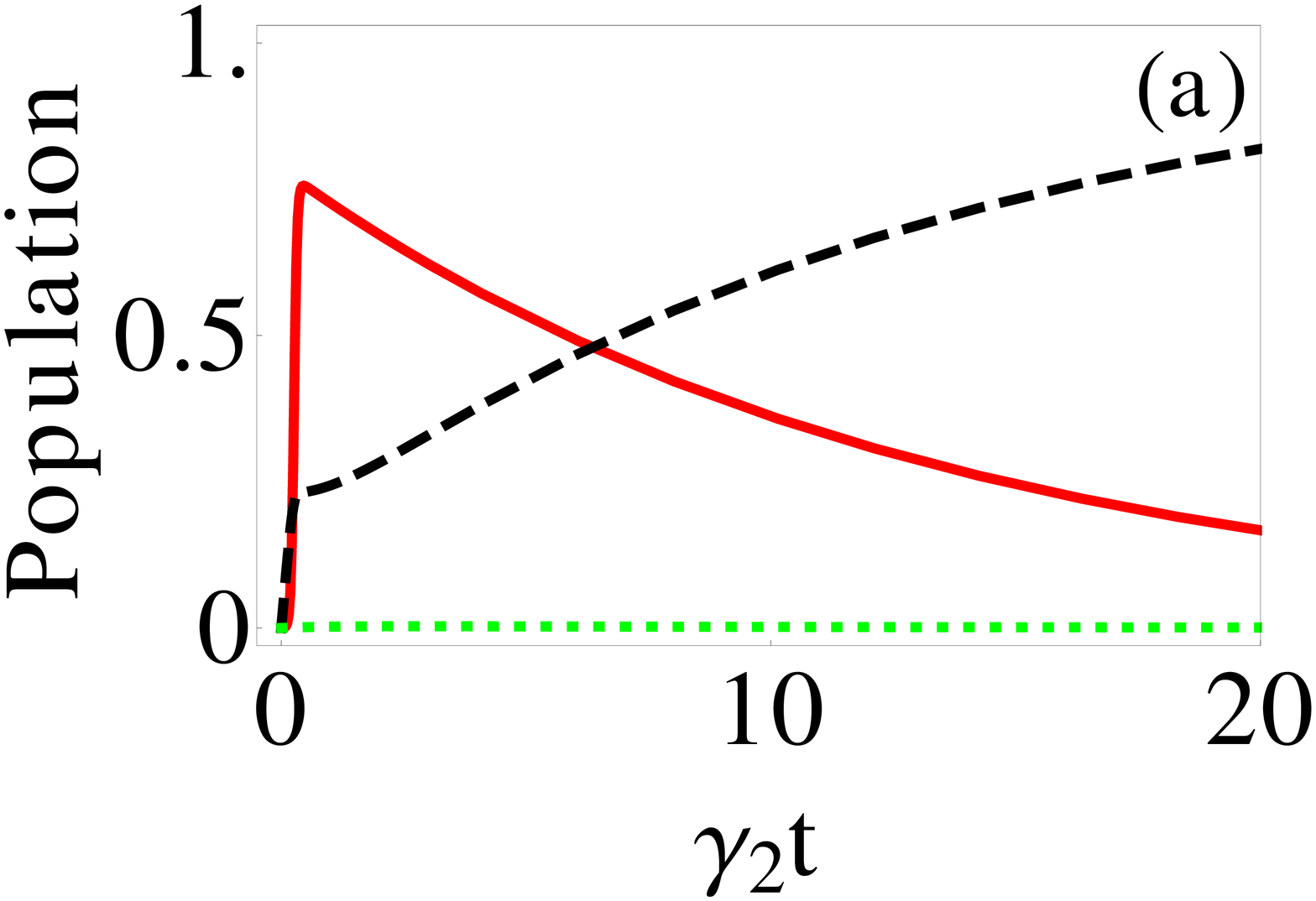}
\hspace{-5mm}
\includegraphics[width=4.35cm,height=3.3cm]{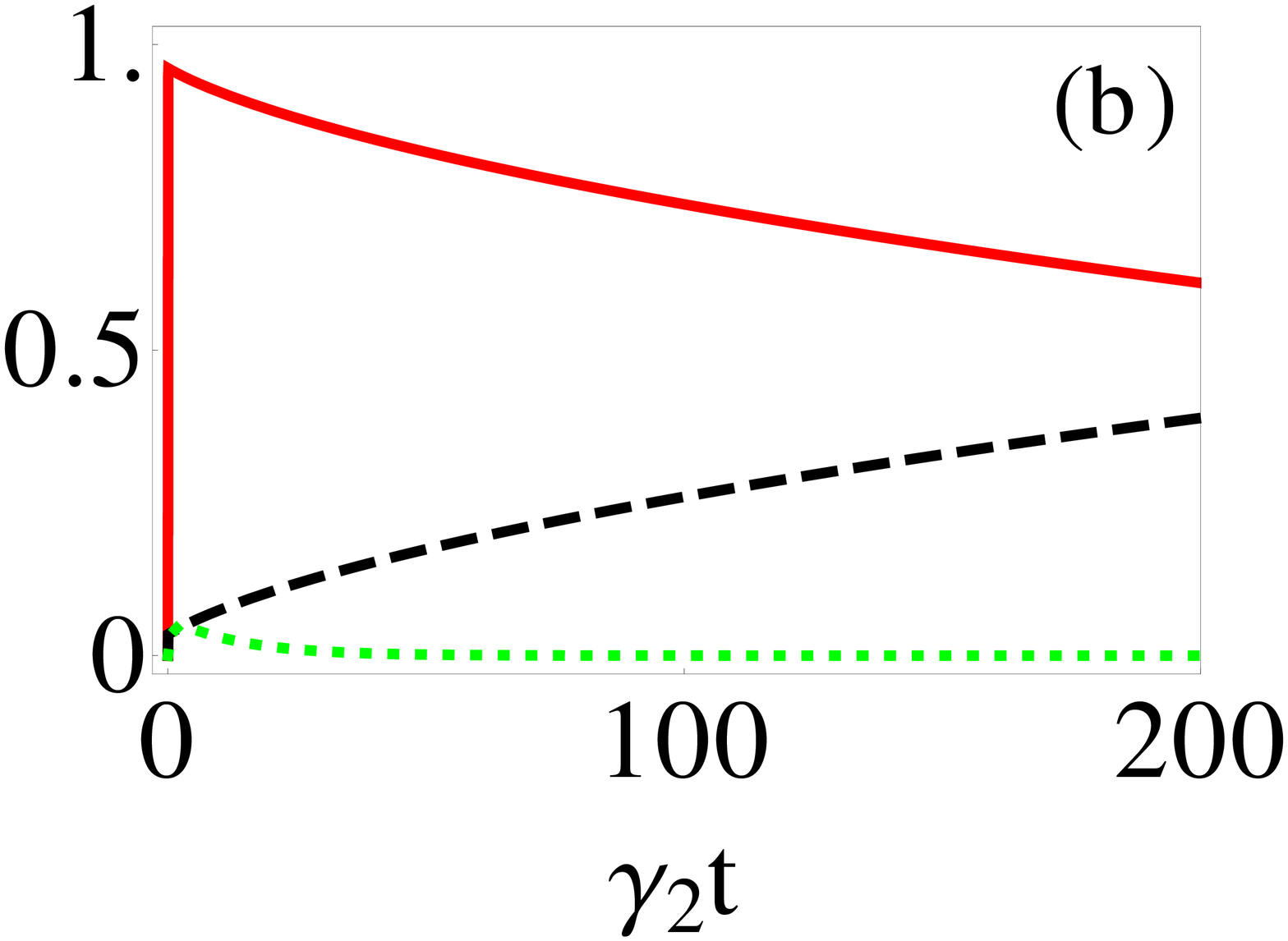}
\caption{\label{fig-3}(color online) The time evolution of the population in the excited $|2P\rangle$ state (red solid curves)
for the collectively interacting Ps atoms and excited by a resonant laser field with a Rabi frequency $\Omega=500\gamma_2$.
(a) the collectivity occurs only on the $|0\rangle \to |1\rangle$ atomic transition;
(b) the Dicke model of Ps atoms additionally coupled by an optical cavity on the transition
$|0\rangle \leftrightarrow |1\rangle$. The green dotted lines are the corresponding results for an
independent Ps atom. The dashed black curve describes the annihilated atoms.}
\end{figure}

To increase the Ps lifetime, we employ a coherent resonant circularly polarized driving laser field which drives the atoms between the $1S$ and $3D$ states. The density of the Ps ensemble ($\rho$) is chosen in a way that the radiative transition $3D \rightarrow 2P$  is collective,  but, at the same time, the radiators are independent on the transition $2P \rightarrow 1S$: $\lambda_{01}^{-3} \ll\rho\ll \lambda_{12}^{-3}$. 
Then, due to  the constructive interatomic interference, the Ps atoms rapidly decay to the $2P$ state cooperatively and 
are trapped in the excited $2P$ state for a much longer time than independent Ps atoms and this results in a significant increase of the annihilation lifetime. 
Without collective radiative transitions, only little population will be collected in the $2P$ state. In fact,
the spontaneous decay on the $2P \rightarrow 1S$ transition is much faster than that of
$3D \rightarrow 2P$ as $\gamma_0 < \gamma_1$ and
there will not be enough time for accumulating the population in the excited $2P$ state before decaying
to the $1S$ ground state. It can be seen from Fig.~\ref{fig-3}(a) that the population of the $2P$ state
is almost zero for an independent Ps atom even when coupled by a strong laser field with $\Omega=500\gamma_2$.
In contrast to this, when collective effects in the transition $3D \rightarrow 2P$ enter into play,  enhancing the transition probability while the radiators are independent on the transition $2P \rightarrow 1S$, 
almost 80$\%$ of the population can be pumped into the excited $2P$ state with a longer lifetime, 
see Fig.~\ref{fig-3}(a). Accordingly, the annihilation lifetime is  extended about 20 times.

There is a possibility for further increasing the trapping capability of our scheme at a fixed density of Ps atoms. 
We propose the following strategy for optimized trapping. We choose a regime when collectivity is present on both the dipole-allowed transitions $3D \rightarrow 2P$  and $2P \rightarrow 1S$: $\rho\gg \lambda_{12}^{-3}$. Then, the population will be collected in the $2P$-state if the collective transition rate $3D \rightarrow 2P$ is larger than the one for the $2P \rightarrow 1S$ transition.
The collective transition rate $3D \rightarrow 2P$ is proportional to $\gamma_0$ and the increase of the latter will increase in total the impact of the collective effects on the trapping.
It is well-known that the rate of spontaneous emission depends partly on the surrounding environment, in particular, 
on the vacuum mode density, which can be modified, e.g., via an optical cavity, similar to the Purcell effect \cite{Purcell1,Purcell2}
(confirmed also experimentally~\cite{Purcell-exp}). Therefore, in order to enhance the spontaneous decay rate $\gamma_0$, we place the ensemble of Ps atoms in a cavity (see Fig.~\ref{fig-1}b), whose fundamental mode frequency is resonant with the $3D \leftrightarrow 2P$ transition frequency. Due to the resonant coupling, the spontaneous decay rate $\gamma_0$ changes into $\gamma^{\prime}_0=g^2/\kappa$~\cite{Macovei,Knight}, where $g$ is the atom-cavity coupling strength and $\kappa$ is
the cavity damping rate. A cavity-induced enhancement occurs if $g^2/\kappa$ is larger than the free space decay rate $\gamma_0$. Using an enhanced decay rate $\gamma_0 =1.6 \gamma_1$,
we solve the master equation Eq.~(\ref{meq}) for the case when collectivity is present
on both dipole-allowed transitions. The time evolution of population in the excited $2P$ state is represented in Fig.~\ref{fig-3}(b) for an ensemble of $N=10^4$ Ps atoms excited by the resonant laser field with $\Omega=500\gamma_2$. In the case of an independent Ps atom, the resonant laser field drives the population in the excited $3D$ state and only 5$\%$ of population remains in the excited $2P$ state at the very beginning due to the fast decay from the $3D$ state to the $2P$ state. Nevertheless, for an ensemble of collectively
interacting Ps atoms, 
the population is almost completely transferred to the excited $2P$ state at the very beginning. For higher interaction times, the population in the $2P$ state decreases quite slowly but steadily, and eventually reaches zero at very long times $\gamma_2t \gg 100$. As a result, the annihilation lifetime is extended by about a factor of more than $200$, which is much larger than in previous works ~\cite{Pazdzerskii,Faisal,Mittleman-1986,Mittleman}.

In summary, we have shown that the cooperative spontaneous emission available at high density of a Ps ensemble 
can be employed to manipulate the annihilation dynamics. In particular, the collective effects for a Ps ensemble, 
coupled simultaneously with a laser field and an optical cavity are shown to allow for the trapping of atoms in 
the excited $p$-state and, in this way, for significantly increasing the Ps annihilation lifetime and facilitating Ps 
BEC formation.




\begin{thebibliography}{30}
\bibitem{Rich} A. Rich, Rev. Mod. Phys. {\bf 53}, 127 (1981).

\bibitem{Mills-1994} P. M. Platzman and A. P. Mills, Jr., Phys. Rev. B {\bf 49}, 454 (1994).

\bibitem{Mills-2010} D. B. Cassidy, V. E. Meligne, and A. P. Mills, Jr., Phys. Rev. Lett. {\bf 104}, 173401 (2010).

\bibitem{Cassidy} D. B. Cassidy, \textit{et al.}
Phys. Rev. Lett. {\bf 106}, 023401 (2011).

\bibitem{gamma-laser} A. Loeb and S. Eliezer, Laser and Particle Beams {\bf 4}, 577 (1986).

\bibitem{gamma-laser2} E. P. Liang and C. D. Dermer, Opt. Commun. 65, 419 (1988).

\bibitem{Mills-2002} A. P. Mills, Jr., Nucl. Instrum. Meth. Phys. Res. B {\bf 192},  107 (2002); A. P. Mills, Jr., D. B. Cassidy, and R. G. Greaves, Mat. Sci. Forum \textbf{445}-\textbf{446}, 424 (2004).

\bibitem{anti-H} B. I. Deutch, A. S. Jensen, A. Miranda and G. C. Oades, Proc. 1st Workshop on Antimatter Physics at Low Energies, Fermilab (FNAL, Batavia, IL,1986) p. 371; J. W. Humberston, M. Charlton, F. M. Jacobsen, B. I. Deutch, J. Phys. B {\bf 20}, L25 (1987).

\bibitem{Mills}
D. B. Cassidy,  \textit{et al.}, Phys. Rev. Lett. \textbf{107}, 033401 (2011); \textit{ibid.} \textbf{106}, 173401 (2011); \textit{ibid.} \textbf{106}, 133401 (2011); S. Mariazzi, P. Bettotti, and R. S. Brusa, \textit{ibid.} {\bf 104}, 243401 (2010); D. B. Cassidy, \textit{et al.}  Phys. Rev. A \textbf{81}, 012715 (2010);  C. M. Surko and R. G. Greaves, Phys. Plasmas \textbf{11}, 2333 (2004); P. Perez and A. Rosowsky, Nucl. Instr. Meth.  Phys. Res. A \textbf{532}, 523 (2004); R. G. Greaves and C. M. Surko, Phys. Rev. Lett. \textbf{85}, 1883 (2000); J. Estrada \textit{et al.}, \textit{ibid.} \textbf{84}, 859 (2000).



\bibitem{Mills_Nature} D. B. Cassidy and A. P. Mills, Jr., Nature \textbf{449}, 195 (2007).


\bibitem{Mills-1975} A. P. Mills, Jr., S. Berko, and K. F. Canter, Phys. Rev. Lett. {\bf 34}, 1541 (1975).

\bibitem{Pazdzerskii} V. A. Pazdzerskii, Soviet Physics Journal {\bf 20}, 963 (1977).

\bibitem{Faisal} F. H. M. Faisal and P. S. Ray, J. Phys. B: At. Mol. Phys. {\bf 14}, L715 (1981).

\bibitem{Mittleman-1986} M. H. Mittleman, Phys. Rev. A {\bf 33}, 2840 (1986).

\bibitem{Mittleman} A. Karlson and M. H. Mittleman, J. Phys. B {\bf 29}, 4609 (1996).

\bibitem{Lambropoulos} L. B. Madsen, L. A. A. Nikolopoulos and P. Lambropoulos, J. Phys. B \textbf{32}, L425 (1999).

\bibitem{Ziock} K. P. Ziock, \textit{et al.}, Phys. Rev. Lett. {\bf 64}, 2366 (1990).

\bibitem{Cassidy_2005}
D. B. Cassidy, et al. Phys. Rev. Lett. \textbf{95}, 195006 (2005).

\bibitem{Agarwal} G. S. Agarwal, {\it Quantum Statistical Theories of Spontaneous Emission and their Relation to other Approaches}, (Springer, Berlin, 1974).

\bibitem{Ficek} Z. Ficek and S. Swain, {\it Quantum Interference and Coherence: Theory and Experiments}, (Springer, Berlin, 2005).

\bibitem{Macovei} M. Kiffner, M. Macovei, J. Evers, and C. H. Keitel, in {\it Progress in Optics}, edited by E. Wolf (North Holland, Amsterdam, 2010), Vol. {\bf 55}, p. 85.

\bibitem{Gross}M. Gross and S. Haroche, Phys. Rep. {\bf 93}, 301 (1982).

\bibitem{Dicke} R. H. Dicke, Phys. Rev. {\bf 93}, 99 (1954).

\bibitem{Crubellier} A. Crubellier, S. Liberman, and P. Pillet, Opt. Commun. {\bf 33}, 143 (1980).

\bibitem{Pavolini} D. Pavolini, A. Crubellier, P. Pillet, L. Cabaret, and S. Liberman, Phys. Rev. Lett. {\bf 54}, 1917 (1985).

\bibitem{Purcell1} E. M. Purcell, H. C. Torrey, and R. V. Pound, Phys. Rev. {\bf 69}, 37 (1946).

\bibitem{Purcell2} E. M. Purcell, Phys. Rev. {\bf 69}, 681 (1946).

\bibitem{Purcell-exp} P. Goy, J. M. Raimond, M. Gross, and S. Haroche, Phys. Rev. Lett. {\bf 50}, 1903(1983).

\bibitem{Knight} B. M. Garraway and P. L. Knight, Phys. Rev. A {\bf 54}, 3592 (1996).
\end{thebibliography}
\end{document}